\documentstyle[preprint,12pt,aps]{revtex}
\tightenlines
\input epsf
\begin{document}
\title{Analytic computation of the energy levels of a two-dimensional hydrogenic
donor in a constant magnetic field}
\author{V\'{\i}ctor M. Villalba
\thanks{e-mail:villalba@ivic.ivic.ve}, and Ramiro Pino\thanks{e-mail:rpino@ivic.ivic.ve}}
\address{Centro de F\'{\i}sica\\
Instituto Venezolano de Investigaciones Cient\'{\i}ficas, IVIC\\
Apdo 21827, Caracas 1020-A, Venezuela}
\maketitle

\begin{abstract}
We compute the energy levels of a 2D Hydrogen atom when a constant magnetic
field is applied. With the help of a mixed-basis variational method, we calculate the energy eigenvalues of the $%
1S $, $2P^{-}$ and $3D^{-}$ levels. We compare the computed energy spectra
with those obtained via a generalization of the mesh point technique as well
as the shifted $1/N$ method. We show that the variational solutions present
a good behavior in the weak and strong magnetic field regimes.
\end{abstract}


The study of the behavior of  hydrogenlike atoms in magnetic fields has been a
subject of considerable interest in recent years. Stimulated by recent
advances in nanofabrication technology that have made possible to create low
dimensional structures containing one or more electrons: quantum wells,
wires, and dots, a large body of papers\cite{Bastard,Greene,Zhu,Sikorski} (and references therein) has considered the
problem of two-dimensional donors in semiconductor superlattices. The most
commonly studied semiconductor superlattice consists of regions of $GaAs$
which act as wells for the conduction electrons separated by regions of $%
Al_{x}Ga_{1-x}As$ which act as barriers. Quantum wells in the presence of
strong magnetic fields can form quantum dots, therefore the study of donor
levels in $Ga_{1-x}Al_{x}As$ $\ $would be of help in understanding the role
played by impurity ions in low dimensional structures. The computation of
the binding energies as well as the study of exchange and correlation
effects make it necessary to have at our disposal reliable analytic as well as
numerical methods. In this direction, different approaches have been
considered. Besides numeric and perturbation methods \cite
{Whittaker,Duggan,villalba}, we have to mention the two-point Pad\'{e}
approximation \cite{MacDonald,Martin}, as well as the shifted 1/N expansion 
\cite{Mustafa1,Mustafa2}.

The Hamiltonian describing the Coulomb interaction between a conduction
electron and a donor impurity center when a constant magnetic field B is
applied perpendicular to the x-y plane can be written as

\begin{equation}
H=-\nabla ^{2}+\gamma L_{z}-\frac{2}{\rho }+\frac{\gamma ^{2}\rho ^{2}}{4}
\label{Ham}
\end{equation}
where we have chosen the vector potential {\bf A} =$\frac{Br}{2}\hat{e}%
_{\varphi }$ in the symmetric gauge. The coupling constant is 
$\gamma =\epsilon ^{2}\hbar ^{3}B/(ce^{3}m^{\ast 2})$ where $m^{\ast }$ is
the effective mass, $\epsilon$ is  the dielectric constant of the host
material. $L_{z}$ is the angular momentum operator $-i\hbar \partial
/\partial \phi $ with eigenvalue $\hbar m.$The units of energy are given in
terms of the effective Rydberg constant ${\cal R}_{0}^{\ast }=m^{\ast
}e^{4}/2\hbar ^{2}\epsilon ^{2}$ and the effective Bohr radius $a^{\ast
}=\hbar ^{2}\epsilon /m^{\ast }e^{2},$ respectively.

The substitution 
\begin{equation}
\Psi =\frac{e^{im\varphi }}{\sqrt{2\pi }}\frac{u(\rho )}{\sqrt{\rho }}
\end{equation}
reduces the Schr\"odinger equation $H\Psi =E\Psi $ to the following second
order ordinary differential equation

\begin{equation}  \label{ecua}
\left[ -\frac{d^2}{d\rho ^2}+(m^2-\frac 14)\frac 1{\rho ^2}+\frac{\gamma
^2\rho ^2}4-\frac 2\rho +m\gamma -E\right] u(\rho )=0
\end{equation}

It is well known that exact solutions of Eq. (\ref{ecua}) cannot be
expressed in closed form in terms of special functions\cite{Bagrov}. There
are analytical expressions for the energy for particular values of $\gamma $
and $m$\cite{Lozanskii,Taut1,Taut2}. In this paper, with the help of a
mixed-basis variational method, we report a calculation of the $1S,$ $2P^{-}$
and $3D^{-}$ energy levels of a 2D (x-y plane) hydrogenic atom in the
presence of a strong magnetic field in the z direction; we compare our
results with those obtained using the Schwartz \cite{Schwartz} interpolation
technique; as well as the shifted $1/N$ method.

In order to apply the variational method to our problem \cite{Davydov}, we
look for a trial wave function. Since Eq. (\ref{ecua}) reduces to the
Hydrogen atom equation when $\gamma =0,$ we can consider as a basis for $%
\gamma <<1$ the Hydrogen wave functions $\Psi _{H}.$ Since $<\Psi _{H}\left|
H\right| \Psi _{H}>\leq E,$, we obtain an upper bound of the energy for
small values of the parameter $\gamma .$ The solution of Eq. (\ref{ecua})
when $\gamma =0$ is

\begin{equation}
u_{H}(\rho )=D_{m,n}e^{-\rho /(1/2+n_{\rho }+\left| m\right| )}\rho
^{(\left| m\right| +1/2)}L(n_{\rho },2\left| m\right| ,\frac{2\rho }{%
(1/2+n_{\rho }+\left| m\right| )})
\end{equation}
where $D_{m,n}$ is a normalization constant, and L(a,b,x) are the Laguerre
polynomials\cite{Lebedev}. Consequently the energy spectrum in the
zero-field limit takes the form, 
\begin{equation}
E_{H}=-\frac{1}{(1/2+n_{\rho }+\left| m\right| )^{2}}
\end{equation}
Conversely, for large values of $\gamma ,$ a good trial basis is that of the
spherical oscillator. In this case the solution of Eq. (\ref{ecua}) has the
form 
\begin{equation}
u_{Osc}(r)=C_{m,n}e^{-\gamma \rho ^{2}/4}\rho ^{(\left| m\right|
+1/2)}L(n_{\rho },\left| m\right| ,\frac{\gamma }{2}\rho ^{2})
\end{equation}
and, in the high-field limit the energy levels are 
\begin{equation}
E_{Osc}=\gamma (2n_{\rho }+\left| m\right| +m+1)
\end{equation}

If we attempt to apply the variational method using only the hydrogen atom
basis, we will obtain a good agreement with the accurate results for small
values of $\gamma ,$ but this approach fails for large $\gamma $ even if we
consider a basis with many. Analogous situation occurs with the oscillator
basis, for large $\gamma ,$ which converges very slowly for small
values of $\gamma $. \cite{villalba}.

In order to solve this problem, we propose a mixed-basis approach. The idea
is to use as trial function, for any quantum level $n$, a linear combination
of the form 
\begin{equation}
u_{n}=\sum_{i}^{N}(c_{iH}u_{iH}+c_{N+iO}u_{iOsc})
\end{equation}
where $N\geq i\geq n$; $u_{iH}$ and $u_{iOsc}$ are the corresponding
hydrogen and oscillator wave functions associated with the quantum\ level $i$
; $c_{iO\text{ }}$and $c_{iH}$ are constants to be calculated. It is worth
noticing that our mixed basis is not orthogonal under the inner product $%
\int_{0}^{\infty }u_{i}u_{j}d\rho .$ We proceed to minimize the expectation
value $\left\langle u\left| H\right| u\right\rangle $ with the normalization
condition, $\left\langle u|u\right\rangle =1$

After performing a variation on the basis coefficients $c_{i},$ we reduce our
problem to that of solving the matrix equation

\begin{equation}
\left[ \left\langle u_{i}\left| H\right| u_{j}\right\rangle -\lambda
\left\langle u_{i}|u_{j}\right\rangle \right] c_{j}=0  \label{var}
\end{equation}
after substituting the Hamiltonian (\ref{Ham}) into (\ref{var}), where
the lowest value of $\lambda $ will be the energy of the level. When $j\leq
3 $, we can analytically compute the energy eigenvalues $\lambda$ and
eigenvectors $c_{j}$. The advantage of this approach is twofold. First, we
have that the eigenvalues satisfy the inequality $\lambda \leq E$ and
therefore we have a lower bound for our energy levels. Second, we obtain a
relatively simple expression for the normalized eigenfunctions.

In this paper we choose to work with a three term mixed-variational basis.
In this case we can solve the resulting third-order algebraic equation (\ref
{var}) with the help of the Cardano method.

In order to compute the binding energy for the ground $1S$ state, following
the above proposed scheme, we have two possible trial functions 
\begin{equation}
u_{1S}=c_{1}u_{1SH}+c_{2}u_{1SO}+c_{3}u_{2SO}\quad (mix12)
\end{equation}
and 
\begin{equation}
u_{1S}=c_{1}u_{1SH}+c_{2}u_{2SH}+c_{3}u_{1SO}\quad (mix21)
\end{equation}
It is worth mentioning that our three-term bases have the same angular
dependence of the eigenfunction to be approximated. Analogously, 
\begin{equation}
u_{2P^{-}}=c_{1}u_{2P^{-}H}+c_{2}u_{2P^{-}O}+c_{3}u_{3P^{-}O}\hspace{0.1cm}
(mix12),\hspace{0.2cm}
u_{2P^{-}}=c_{1}u_{2P^{-}H}+c_{2}u_{3P^{-}H}+c_{3}u_{2P^{-}O}\hspace{0.1cm} (mix21)
\end{equation}
and 
\begin{equation}
u_{3D^{-}}=c_{1}u_{3D^{-}H}+c_{2}u_{3D^{-}O}+c_{3}u_{4D^{-}O}\hspace{0.1cm}
(mix12),\hspace{0.1cm}
u_{3D^{-}}=c_{1}u_{3D^{-}H}+c_{2}u_{4D^{-}H}+c_{3}u_{3D^{-}O}\hspace{0.2cm} (mix21)
\end{equation}

The numerical computations of the energy spectra associated with Eq. (\ref
{ecua}) will be carried out with the help of the Schwartz method \cite
{Schwartz} This method gives highly accurate results given a thoughtful
choice of the reference function. For Eq. (\ref{ecua}) we chose as the
interpolation function 
\begin{equation}
f(\rho )=\sum_{m}f_{m}\frac{u(\rho )}{(\rho -r_{m})a_{m}}
\end{equation}
with $u(\rho )=\sin [\pi (\rho /h)^{1/2}]$ , $r_{m}$ is a zero of $u(\rho )$%
, $a_{m}$ is a zero of its derivative, and $h$ is the step of the quadratic
mesh. The use of this scheme on Eq. (\ref{ecua}) leads to an algebraic
eigenvalue problem, giving as result a non-symmetric matrix to be
diagonalized in order to obtain the energy values.

Here, as an illustration of the mixed-basis method, we present two different
expansions of the $1S$, $2P^-$, and $3D^-$ of the 2D Hydrogen Hamiltonian (\ref{Ham}). We
plot the energy against $\gamma ^{\prime }=\gamma /(\gamma +1)$ as the
horizontal scale. 
\begin{figure}[tbp]
\centerline{\epsffile{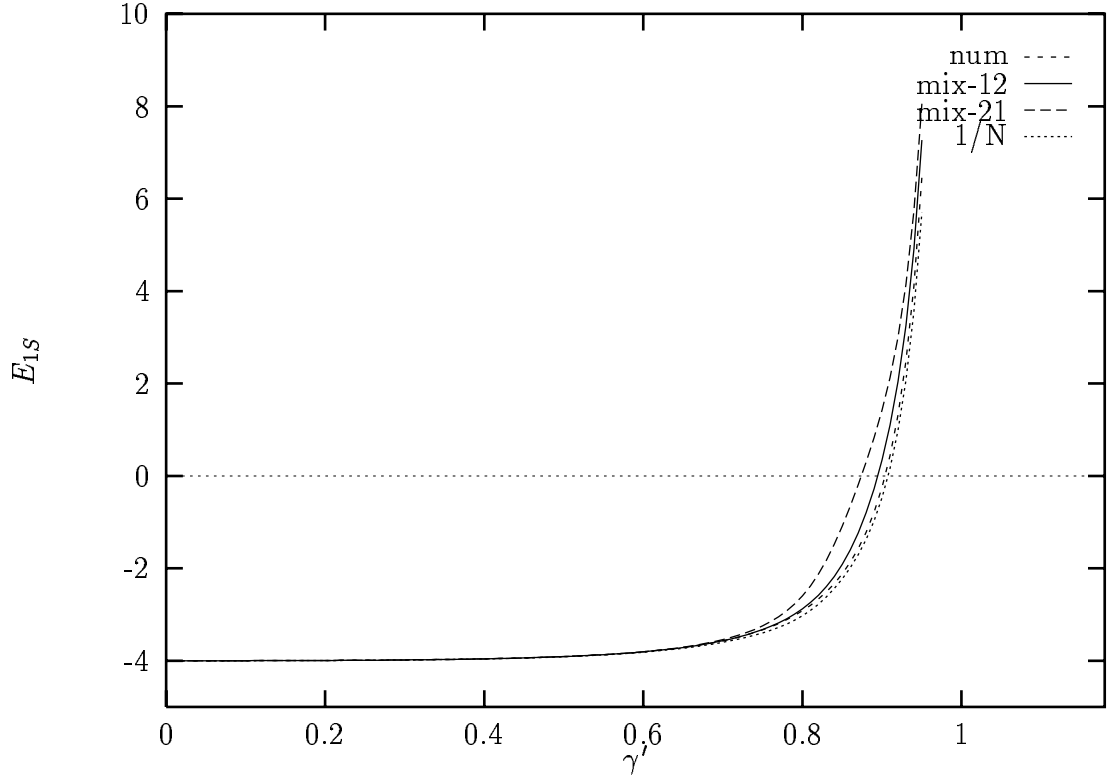}}
\caption{Energy of the $1S$ state as a function of $\protect\gamma^{\prime
} $. The light broken line is obtained by numerical methods; the dashed
line corresponds to the mix21 basis, (2S, 3S Hydrogen bases and 1S oscillator
wavefunction). The solid line is obtained by using the mix12 basis (1S, 2S
oscillator bases and the 1S Hydrogen wavefunction).The dotted line is
obtained with the help of the shifted $1/N$ method}.
\end{figure}

\begin{figure}[tbp]
\centerline{\epsffile{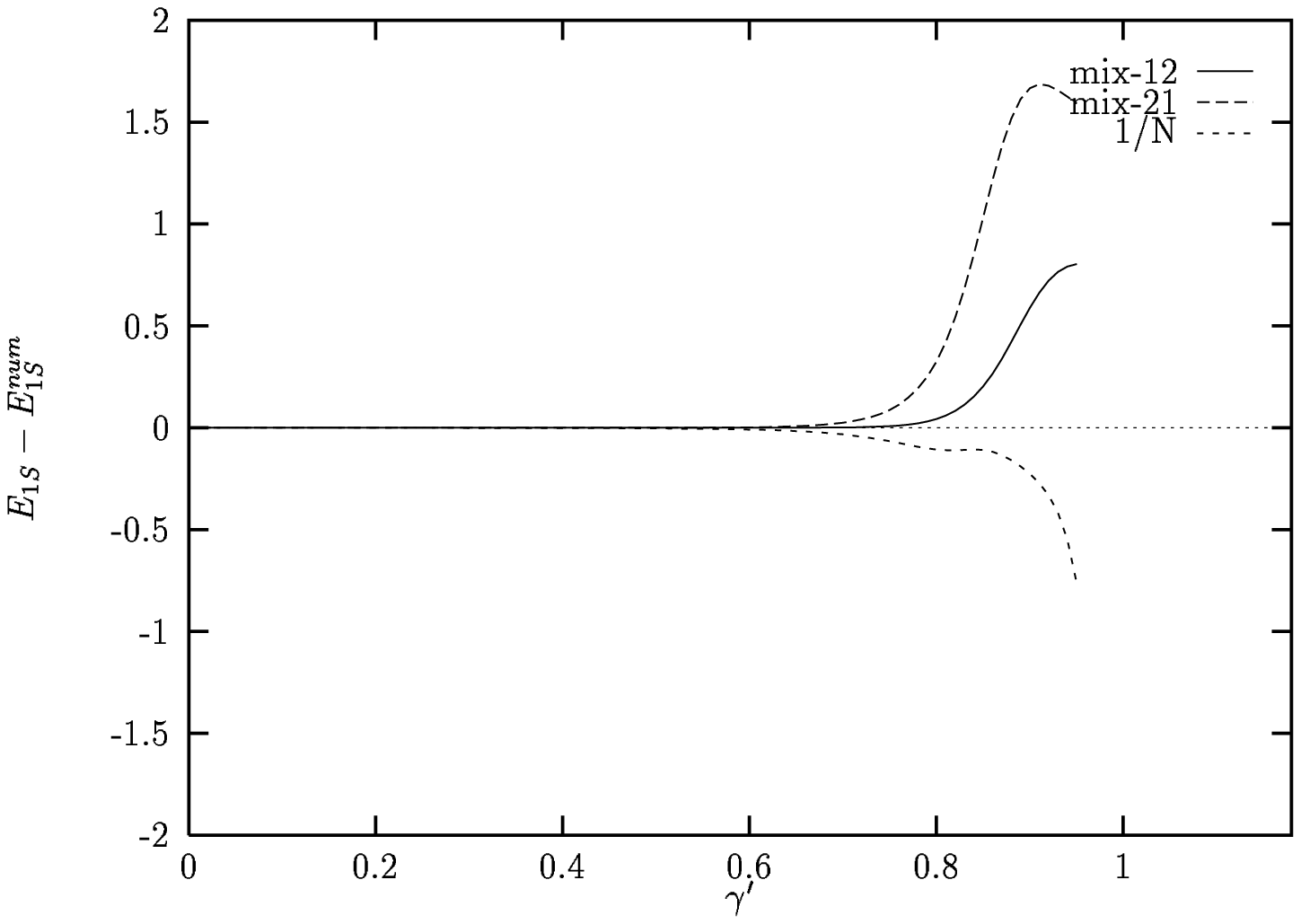}}
\caption{The figure shows the difference between the numeric result for the $1S$
energy spectrum and the energy values computed with the help of the
mix12 basis (solid line), mix21 basis (heavy dashed line),
and the shifted $1/N$ method (light dashed line).}
\end{figure}

\begin{figure}[tbp]
\centerline{\epsffile{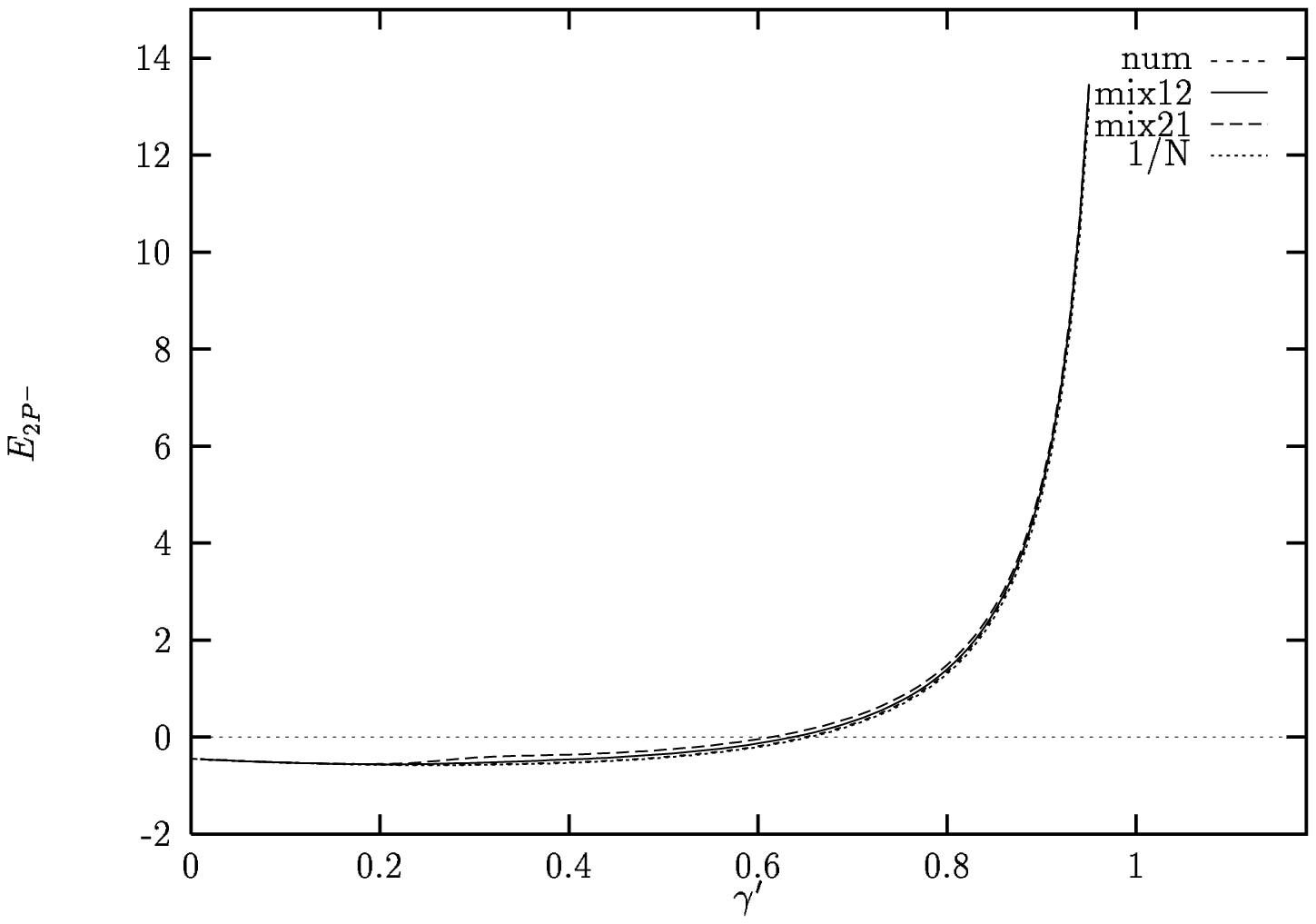}}
\caption{Energy of the $2P^-$ state as a function of $\protect\gamma^{\prime}$
The light broken line is obtained by numerical methods; the dashed 
line corresponds to the mix21 basis,($2P^-$, $3P^-$ Hydrogen bases and $2P^-$ oscillator 
wavefunction). The solid line is obtained by using the mix12 basis ($2P^-$, $3P^-$ 
oscillator bases and the $2P^-$ Hydrogen wavefunction).The dotted line is
obtained with the help of the shifted $1/N$ method.}
\end{figure}

\begin{figure}[tbp]
\centerline{\epsffile{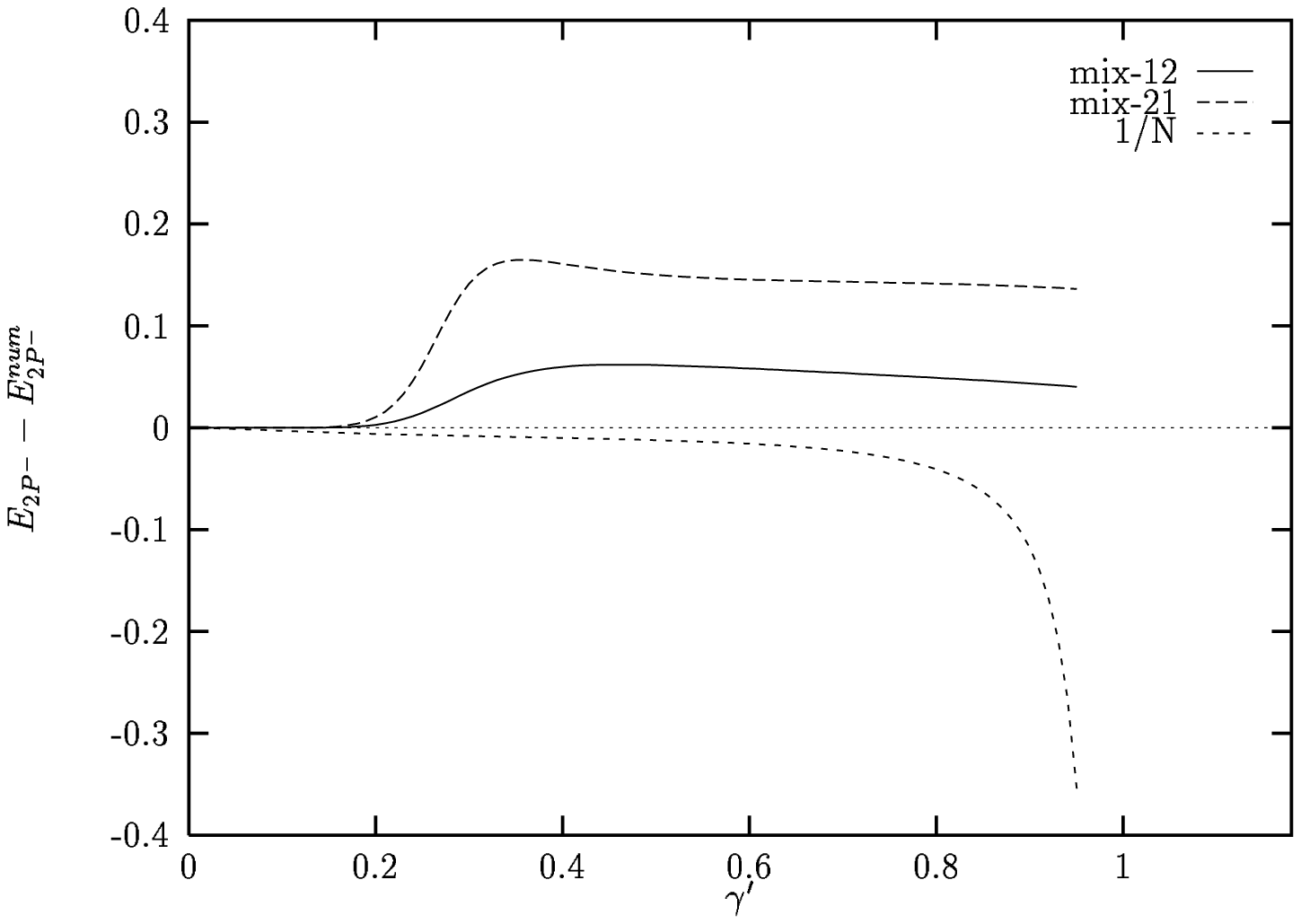}}
\caption{The figure shows the difference between the numeric result for the $2P^-$ 
energy spectrum and the energy values computed with the help of the
mix12 basis (solid line), mix21 basis (heavy dashed line),
and the shifted $1/N$ method (light dashed line).}
\end{figure}

\begin{figure}[tbp]
\centerline{\epsffile{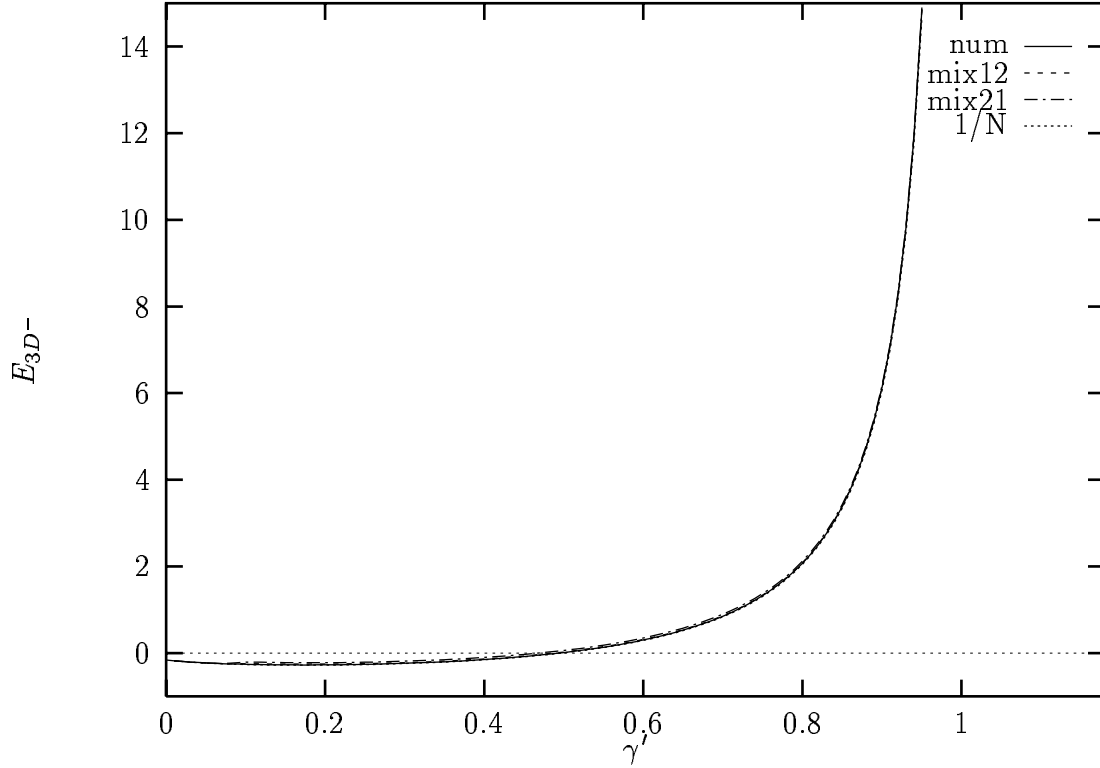}}
\caption{Energy of the $3D^{-}$ state as a function of $\protect\gamma^{\prime }$. The solid line is obtained by numerical methods; the heavy dashed line 
corresponds to the mix21 basis ($3D^{-}$,$4D^{-}$ Hydrogen bases and $3D^{-}$ 
oscillator wavefunction). The light dashed line is obtained by using the mix12 basis ($3D^{-},4D^{-}$ oscillator bases and the $3D^{-}$ Hydrogen wavefunction). The dotted line is obtained with the help of the shifted $1/N$ method.}
\end{figure}

\begin{figure}[tbp]
\centerline{\epsffile{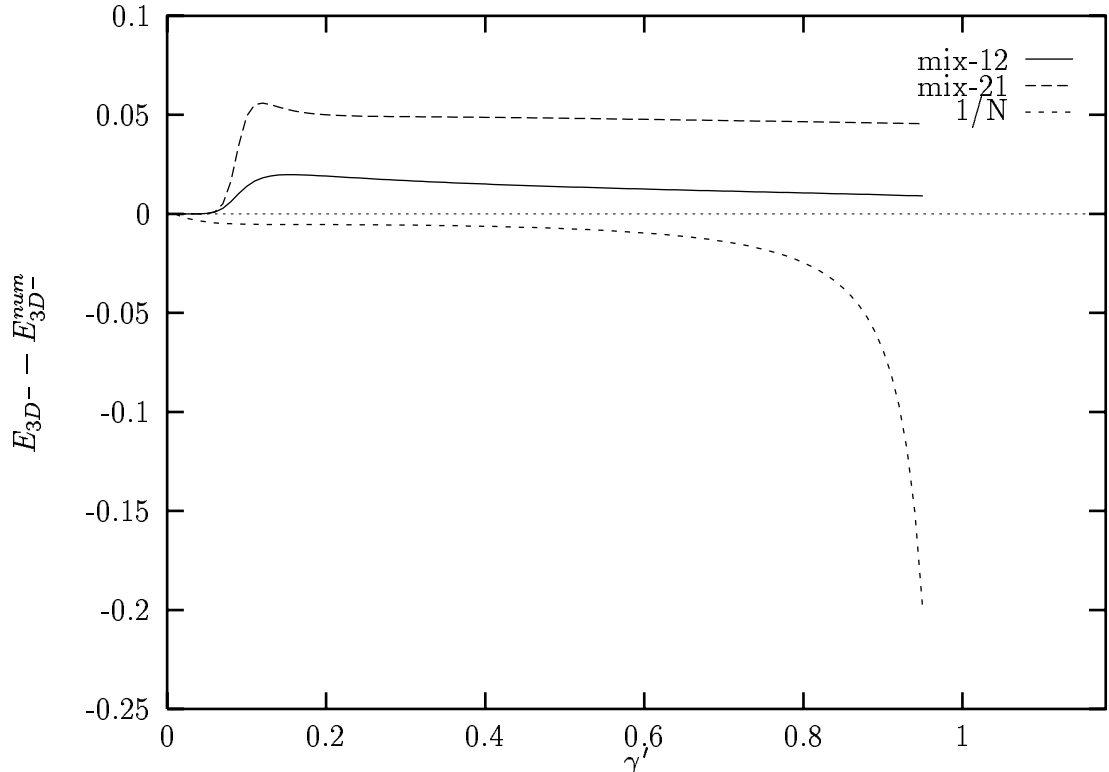}}
\caption{The figure shows the difference between the numeric result for the $3D^-$ 
energy spectrum and the energy values computed with the help of the
mix12 basis (solid line), mix21 basis (heavy dashed line),
and the shifted $1/N$ method (light dashed line).}
\end{figure}

It is easy to see that even for a two term mixed basis a good fit is
obtained in the weak and strong field regimes. One of the basis gives a
reasonably good fitting in the intermediate region. A better fit is obtained
with the help of the mix12 bases. We also have that for the $
2P^{-}$ and $3D^{-}$ states the mixed-basis variational approach gives very
good results. Figures 1, 3, and 5 compare the variational results with those
obtained numerically and with the help of the shifted $1/N$ method.

It is difficult to decide on the evidence of Figs. 1, 3, and 5 which technique is better suited to the computation of the hydrogen energy levels. Figs. 2, 4, and 6 show that the shifted $1/N$ method always gives results below the accurate energy levels. Although Figs. 1, 3, and 5 do not show mayor changes between the variational and shifted $1/N$ method, the difference plots Figs. 2, 4, and 6 clearly show the improved performance of the mixed (mix12) variational method, especially for large values of $\gamma$. 

It would be interesting to apply the mixed-basis technique for the 2D
Hydrogen problem when relativistic effects are not negligible. This will be
the object of a forthcoming publication.

\acknowledgments We thank Dr. Juan Rivero for helpful discussions. This work
was supported by CONICIT under project 96000061.

\end{document}